\documentclass[preprint, 12pt, authoryear,colorlinks=true,linkcolor=black, citecolor=blue, urlcolor=blue]{elsarticle}

\usepackage[section]{placeins}
\usepackage[utf8]{inputenc}
\usepackage{multirow}
\usepackage{mathtools}
\usepackage{amssymb}
\usepackage{amsmath}
\usepackage{bbold}
\usepackage{graphicx}
\usepackage{lineno}
\usepackage{float}
\usepackage[top=1.5in, bottom=1.5in]{geometry}
\RequirePackage{doi}
\usepackage{siunitx}

\begin{document}
\begin{frontmatter}
\title{Eccentricities and the Stability of Closely-Spaced Five-Planet Systems}

\author[Pierre]{Pierre Gratia\corref{cor1}}
\ead{pierregratia@gmail.edu}
\cortext[cor1]{Corresponding author}
 \address[Pierre]{CIERA and Department of Physics and Astronomy, Northwestern University, Evanston, IL 60208}

\author[0000-0001-6513-1659]{Jack J. Lissauer}
\ead{jack.lissauer@nasa.gov}
\address[Jack]{Space Science \& Astrobiology Division, MS 245-3, NASA Ames Research Center, Moffett Field, CA 94035}

\journal{Icarus}

\begin{abstract}
Observations of exoplanets have revealed that systems with planets on closely-spaced orbits are common, which motivates the question ``How closely can planets orbit to one another and still be dynamically-stable for very long times?". To address this question, we investigate the stability of idealized planetary systems consisting of five planets, each equal in  mass to the Earth, orbiting a one solar mass star. All planets orbit in the same plane and in the same direction, and the planets are uniformly spaced in units of mutual Hill Sphere radii. 
Most of the systems that we integrate begin with one or more planets  on eccentric orbits, with eccentricities $e$ as large as $e= 0.05$ being considered. For a given initial orbital separation, larger initial eccentricity of a single planet generally leads to shorter system lifetime,  with little systematic dependence  of which planet is initially on an eccentric orbit. The approximate trend of instability times increasing exponentially with initial orbital separation of the planets found previously for planets with initially circular orbits is also present for systems with initially eccentric orbits.  Mean motion resonances also tend to destabilize these systems, although the reductions in system lifetimes are not as large as for initially circular orbits.  Systems with all planets having initial $e= 0.05$ and aligned periapse angles typically survive far longer than systems with the same spacing in initial semi-major axis and one planet with $e= 0.05$, but they have slightly shorter lifetimes than those with planets initially on circular orbits.
\end{abstract}
\end{frontmatter}

\section{Introduction}

NASA's \textit{Kepler} mission has discovered hundreds of multiple planet systems (\cite{2014ApJ...790..146F}; \cite{2014ApJ...784...45R}).
Many of these multi-planet systems are quite tightly packed, where adjacent planets orbit much closer to one another than do those  in the Solar System (\cite{lissauer2011architecture}). 
For instance, the Kepler-11 system harbors six known planets, five of which have orbital periods between 10 and 47 days  (\cite{lissauer2011closely}). Studying the orbital stability of such tightly-packed planetary systems provides clues to how they form and
how long they survive. The present study can be seen as an extension of both \cite{OBERTAS201752} (henceforth OVT17), who investigated the stability of closely-spaced five-planet systems with the same masses and relative spacings as we use herein and all planets on initially circular orbits; and of \cite{2009Icar..201..381S}, who simulated far  fewer simulations, but include a few systems with both initial eccentricity and inclination. Here, we expand on those findings and look at systems for which one planet starts out with a nonzero initial eccentricity, at a $10$ times higher in initial orbital resolution than \cite{2009Icar..201..381S}. We also consider various systems in which all planets begin on eccentric orbits with $e = 0.05$. 

The stability of systems of two planets has been characterized analytically; see \cite{hill1878researches} and \cite{1993Icar..106..247G}. In contrast, few constraints on the stability of more than two planet systems can be derived analytically. Such an investigation therefore requires extensive numerical integrations, for example \cite{1996Icar..119..261C}; \cite{marzari2002eccentric}; \cite{2009Icar..201..381S}; \cite{0004-637X-807-1-44}; \cite{morrison2016orbital}; \cite{OBERTAS201752}. 

The structure of the paper is as follows. We start by describing the setup of our simulations in Section \ref{sec: methodology}. In Section \ref{sec: initially circular orbits}, the focus is on circular systems and the well-known log-linear relationship between initial planetary separation and time to close encounter, which we reproduce with almost identical regression parameters. We then move on to the core part of this work. In Section \ref{sec: innermost planet eccentric}, we  study systems that start out with the innermost planet eccentric. We consider several different eccentricities and compare the systems' lifetimes. In Section \ref{sec: outer planet eccentric}, we perform similar integrations, giving the outermost planet nonzero eccentricity, and in Section \ref{sec: one intermediate planet eccentric}, we conduct a more limited study with one of the intermediate planets given  nonzero eccentricity. Section \ref{sec: all planets eccentric and aligned orbits} considers systems with all planets initially having eccentricity of 0.05; we study systems in which all planets initially have the same longitude of periapse, as well as systems in which the longitudes of periapse are chosen randomly. We compare the coefficients of the log-linear lifetime relationships for many batches of runs in Section \ref{sec:fits}. Our conclusions are summarized in Section \ref{sec: conclusion}.

\section{Methodology}\label{sec: methodology}
Our systems consist of five identical, 1 M$_\oplus$ (Earth-mass) planets, orbiting a 1 M$_\odot$ (solar mass) star in the same direction, and initially separated by the same multiple of their mutual Hill radii (equation \ref{eq: mutual Hill radius}). We restrict ourselves to a two dimensional study by setting all inclinations to zero.  We investigate the effects of nonzero initial eccentricities on these five-planet systems. 

 In what follows, time is measured in units of the initial orbital period of the innermost planet, i.e., Earth-years if the innermost planet begins at 1 AU. 
Our simulations were performed with REBOUND, the N-body simulation code by \cite{rein2012rebound}. In particular, we use the symplectic Wisdom-Holman integrator (WHFast) provided by REBOUND (\cite{rein2015whfast}). WHFast is fast and unbiased, but ill-suited for resolving collisions and close encounters. Thus, we stop a simulation as soon as a close encounter occurs, and compare lifetimes computed in this manner.  \cite{rice2018survival} investigated similar but shorter-lived systems, and followed their evolution beyond their close encounter time (for systems up to and including initial separations of eight mutual Hill radii), while we focus more closely on long-lived systems and stopped our simulations at the time of close encounter, including systems separated by initial distances well beyond eight mutual Hill radii.

\subsection{Mutual Hill radius} \label{mutual_hill_radius}
Before describing the initial conditions, we recall the definition of the Hill radius, which is our unit of distance between initial planetary orbits. The Hill radius of an astronomical object, here a planet, orbiting a much more massive object, here a star, is the radius of the zone in which the planet's gravitational pull is the primary force affecting  motion relative to the planet. For small eccentricities, it can be approximated as 
\begin{align}
R_H \approx a\Bigl(\frac{m}{3M_\star}\Bigr)^{1/3},
\end{align}
where $M_\star$ and $m$ are the masses of the star and planet, respectively, and $a$ is the semi-major axis. The Hill radius is then proportional to the semi-major axis. In this paper,  following convention for this type of simulations, we  use a slightly modified version: the mutual Hill radius between planet $i+1$ and its immediate neighbor orbiting closer to the host star, planet $i$:
\begin{equation} \label{eq: mutual Hill radius}
R_{H_{i,i+1}} \approx \Bigl(\frac{m_{i+1}+m_i}{3M_\star}\Bigr)^{1/3}\Bigl(\frac{a_i+a_{i+1}}{2}\Bigr).
\end{equation}
Here the denominator of the first ratio continues to be the host star's mass, as in OVT17, but \cite{quarles2018long} use the more accurate formula $M_\star+(i-1)M_{\oplus}$.
The initial separations between the planets are expressed in terms of their mutual Hill radii, with the innermost planet placed at $1$ AU.

\subsection{Initial positions and stopping conditions}
We start by stating the initial conditions and describing the parameters used for our simulations. 
\subsubsection{Initial positions}
The initial semi-major axes of the five planets are always chosen to be separated by the same number (not necessarily integer), $\beta$, times their mutual Hill radii (we follow the notation of \cite{2009Icar..201..381S}, while OVT17 use $\Delta$ instead of $\beta$):

\begin{equation} \label{eq: separation}
a_{i+1} - a_i = \beta R_{H_{i,i+1}}.
\end{equation}
From Equations (\ref{eq: mutual Hill radius}) and (\ref{eq: separation}), we have:
\begin{equation} \label{eq: separation2}
a_{i+1} = a_i\Bigl[1+\frac{\beta}{2}\Bigl(\frac{m_i+m_{i+1}}{3M_\star}\Bigr)^{1/3}\Bigr]\Bigl[1-\frac{\beta}{2}\Bigl(\frac{m_i+m_{i+1}}{3M_\star}\Bigr)^{1/3}\Bigr]^{-1}.
\end{equation}

We perform batches of integrations in which we increment  the initial $\beta$  by 0.01 for each new simulation, while keeping everything else fixed at some specified nonzero initial value.  Systems of two planets on initially circular orbits are stable if $\beta > 2\sqrt{3} \approx 3.4641$, so we begin our numerical studies at $\beta =3.47$. Although this critical two-planet separation criterion is  strictly valid only for circular systems \citep{petit2018hill}, our initial eccentricities are small enough that this starting value for $\beta$ is justified -- not least because we are mainly interested in the long-lived systems.

The true longitude is defined as 
\begin{equation}
\theta \equiv \Omega + \omega + \nu,
\end{equation}
where $\Omega$ is the longitude of ascending node, $\omega$ the argument of pericenter, and $\nu$ the true anomaly.
 Since we are working in a co-planar setting, the longitude of ascending node $\Omega$ is zero, and for convenience we also set the argument of peripasis $\omega=0$. This leaves us with the initial true anomaly $\nu$, which is the angle from the pericenter to the planet's position, and the only nonzero parameter, so the true longitude is $\theta = \nu$. In this work, we choose it as $\theta_i = 2\pi i \lambda$ radians, where $i$ is the $i$-th planet starting from the innermost one with $i=1$, and $\lambda = 1 + \sqrt{5} \approx 1.618$ is the golden ratio. This choice avoids any pair of planets starting near conjunction.

Finally, our time step has been chosen to be  0.049281314 years $\approx 18$ days.

\subsubsection{Stopping conditions}
We stopped any given simulation if a close encounter occurred, which we define as the distance between any two objects at the end of any time step being below $0.01$ AU. We did not encounter a case where a planet has been ejected. We also set two more stopping conditions: one, due to limited computational resources, for any given simulation, namely a maximum run time of $10^{10}$ orbits of the innermost planet's initial orbit (i.e., Earth-years, years hereafter); and one for a given batch of simulations running over a $\beta$-range, namely five of them exceeding $2\times 10^9\approx 10^{9.3}$ years. For  systems that did not have a close encounter within ten billion years, we used ten billion years as the lifetime in the statistics, which produces a smaller bias in our fits than ignoring those runs would.

\section{Initially Circular Orbits} \label{sec: initially circular orbits}

\begin{figure}[!htb]
\begin{center}  
\includegraphics[width=0.70\textwidth]{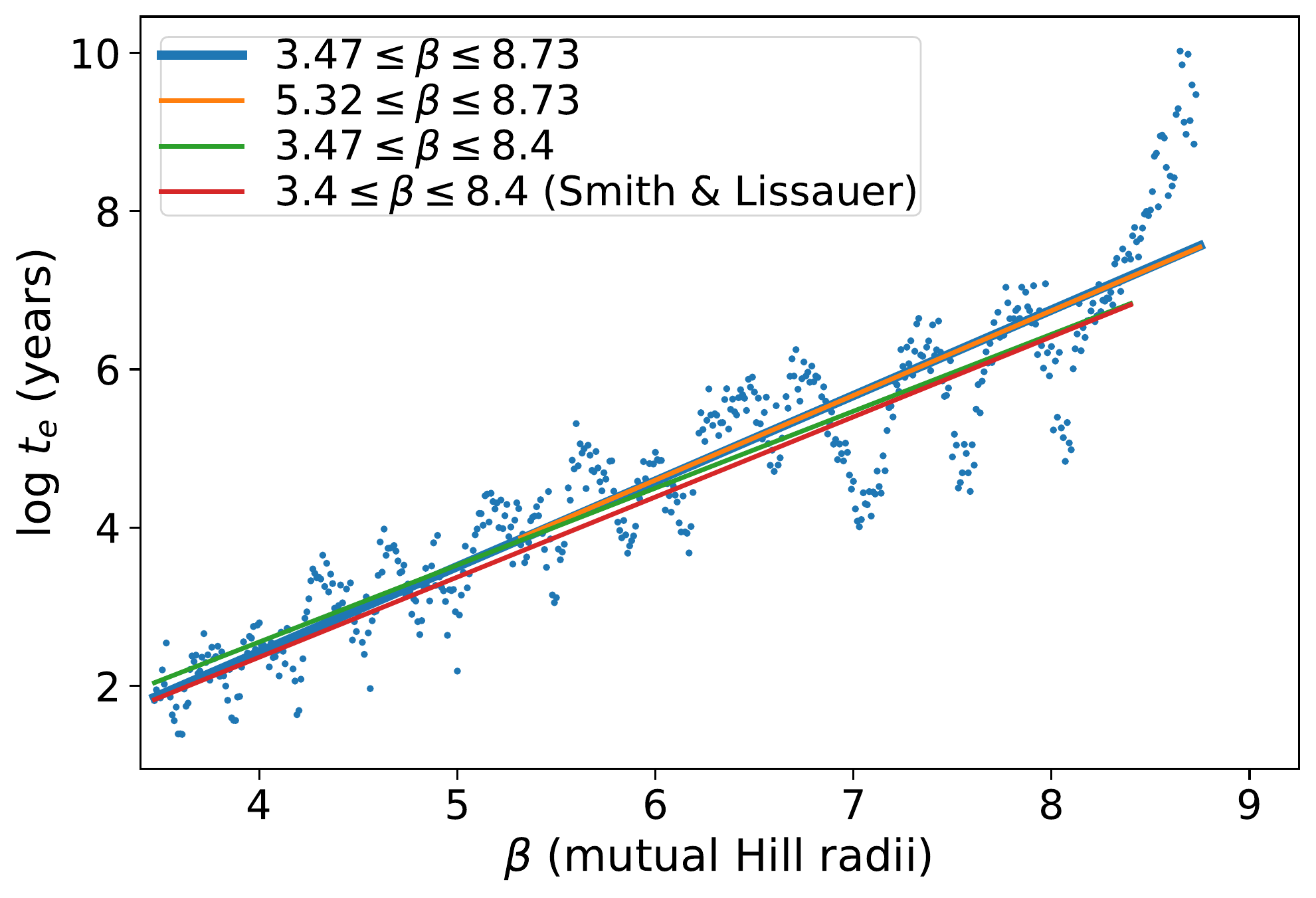} 
\caption{Close encounter times for systems whose planets all start out with zero eccentricity as a function of the separation in initial semi-major axis measured in mutual Hill radii $\beta$. On the vertical axis, $t_e$ stands for ``time to encounter".}
\label{fig:e0_reg}  
\end{center}  
\end{figure}

For the circular case, we ran our simulations in the region $\beta\in[3.47, 8.73]$, the upper bound representing the initial separation of the fifth system that exceeded two billion years before a close encounter occurred. System lifetimes are shown as a function of $\beta$ in Figure \ref{fig:e0_reg}. The general patterns that we find, with lifetimes increasing roughly exponentially with initial separation apart from dips near mean motion resonances, are similar to those noted by SL09 and OVT17.  Our results show somewhat less scatter for similar values of $\beta$ than OVT17, presumably because they performed far more simulations and randomly selected initial planetary longitudes (with the latter choice being especially important for small values of $\beta$).  

In the two subsequent figures, we also added the exponential fits for a selection of $\beta$ ranges. The blue line in Figure \ref{fig:e0_reg} corresponds to the best-fitting exponential dependence of system lifetime as a function of initial orbital separation over the entire interval in $\beta$ that we integrated; the orange line shows the fit for  $5.32 \leq \beta \leq 8.73$, the interval for which we have good stability lifetime values for all sets of runs with the outer four planets beginning on circular orbits; the green line shows the best fit including only systems with $\beta \leq 8.4$; its slope can be compared to previous work by Smith and Lissauer (2009) and OVT17 (see Table \ref{tab:prevfits}). The lines differ only slightly over these $\beta$-ranges, with the largest slope (in orange) belonging to the highest considered lower bound, where $\beta\in [5.32,8.73]$.

\FloatBarrier

The exponential increase in system lifetimes with separation is generally quantified using a log-linear relationship between the close encounter time, $t_e$, and initial separation of the planetary orbital semi-major axes in units of the mutual Hill radius $\beta$: 
\begin{equation}
\log{\left(\frac{t_e}{t_0}\right) = b\beta + c},
\label{eq:power law}
\end{equation}
where $t_0$ is the initial orbital period of the innermost planet (one year), and $b$ and $c$ are constants. As noted in OVT17, however, this exponential fit does not capture all of the variations in close encounter times, which are up to two orders of magnitude off from the regression line. Note also that attempts have been made to fit a power law between initial separation and close encounter time (\cite{zhou2007post}).

We compare our coefficients with those found for the nonzero eccentricity cases in Section \ref{sec: innermost planet eccentric}. A full tabulation of all computed coefficients and their standard errors $\sigma$ can be found in Table \ref{tab:all_fits}. There, however, we considered new parameters $b'$ and $c'$, computed from a new variable $\beta'$ defined by 
\begin{equation}
\beta' \equiv \beta - 2\sqrt{3},
\label{eq:beta prime}
\end{equation}
following \cite{quarles2018long}:

\begin{equation}
\log{\left(\frac{t_e}{t_0}\right) = b'\beta' + c'}.
\label{eq:exp law prime}
\end{equation}

\FloatBarrier

\begin{table}[H]
\begin{center}
    \begin{tabular}{ | l | c | c | l |} \hline
    
    Reference & $b$  & $c$  & Range \\ \hline
 
    Smith and Lissauer (2009) & 1.012 & $-1.686$ & $3.4 \le \beta \le 8.4$  \\ \hline
    Obertas et al. (2017) & 0.951 & $-1.202$ & $2\sqrt{3} < \beta < 8.4$ \\ \hline
    Obertas et al. (2017) & 1.086 & $-1.881$ & $2\sqrt{3} < \beta < 10$ \\ \hline
    This work & $0.972 \pm 0.018$  & $-1.333 \pm 0.107$ & $2\sqrt{3} < \beta \le 8.4$ \\ \hline
    This work & $1.080 \pm 0.019$ & $-1.895 \pm 0.119$ & $2\sqrt{3} < \beta \le 8.73$ \\ \hline


    \end{tabular}
    \caption{
    Comparison of slopes and intercepts of exponential fits to lifetimes of initially circular five planet systems in our simulations with those of previous studies. }
     \label{tab:prevfits}
\end{center}
\end{table}


\section{Innermost planet eccentric} \label{sec: innermost planet eccentric}

\begin{figure}[H]
\begin{center}  
\includegraphics[width=0.70\textwidth]{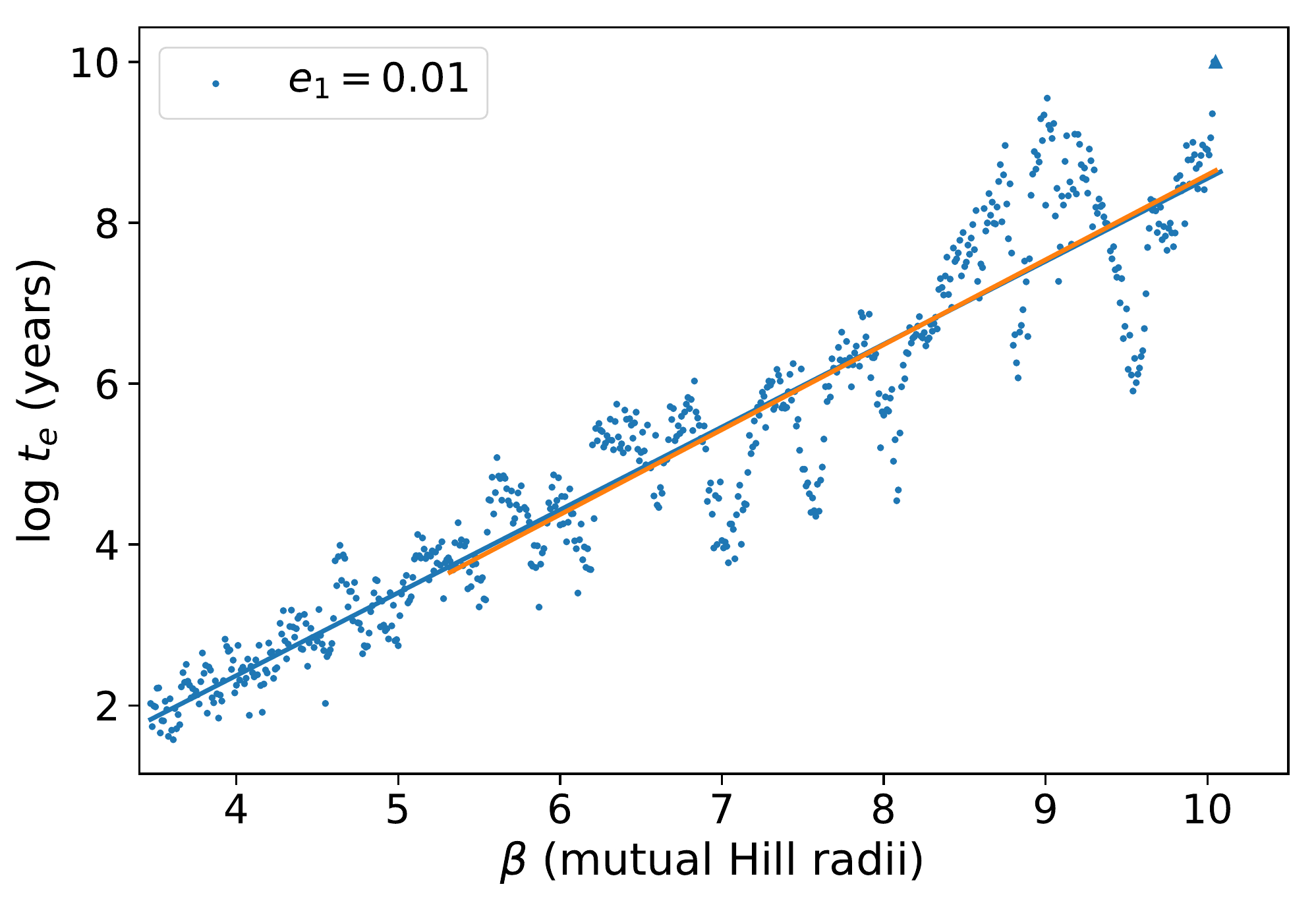} 
\caption{Close encounter times for systems whose innermost planet starts out with eccentricity $e_1=0.01$; the upward-pointing triangle in the upper right represents a system that survived for the entire $10^{10}$ year interval simulated. In blue the corresponding regression line; in orange, the regression line for the $\beta$ interval for which we have values for all batches with the inner planet eccentric and the outer four planets beginning on circular orbits, which is $[5.32, 10.05]$. All slopes and intercept values can be found in Table \ref{tab:all_fits}.}
\label{fig:e001_reg}  
\end{center}  
\end{figure}

We now turn to the investigation of systems with initial conditions analogous to those considered in Section \ref{sec: initially circular orbits}, with the exception that the innermost planet starts out with a nonzero eccentricity.  In our first batch of simulations, we set $e_1=0.01$. The resulting lifetimes are plotted in Figure \ref{fig:e001_reg}. Predictably, compared to the circular case, the close encounter times of those systems tend to be shorter for given $\beta$; hence we needed to go to higher values of $\beta$ to find five systems that survived for $> 2\times10^9$ years.

\begin{figure}[H]
\begin{center}  
\includegraphics[width=0.70\textwidth]{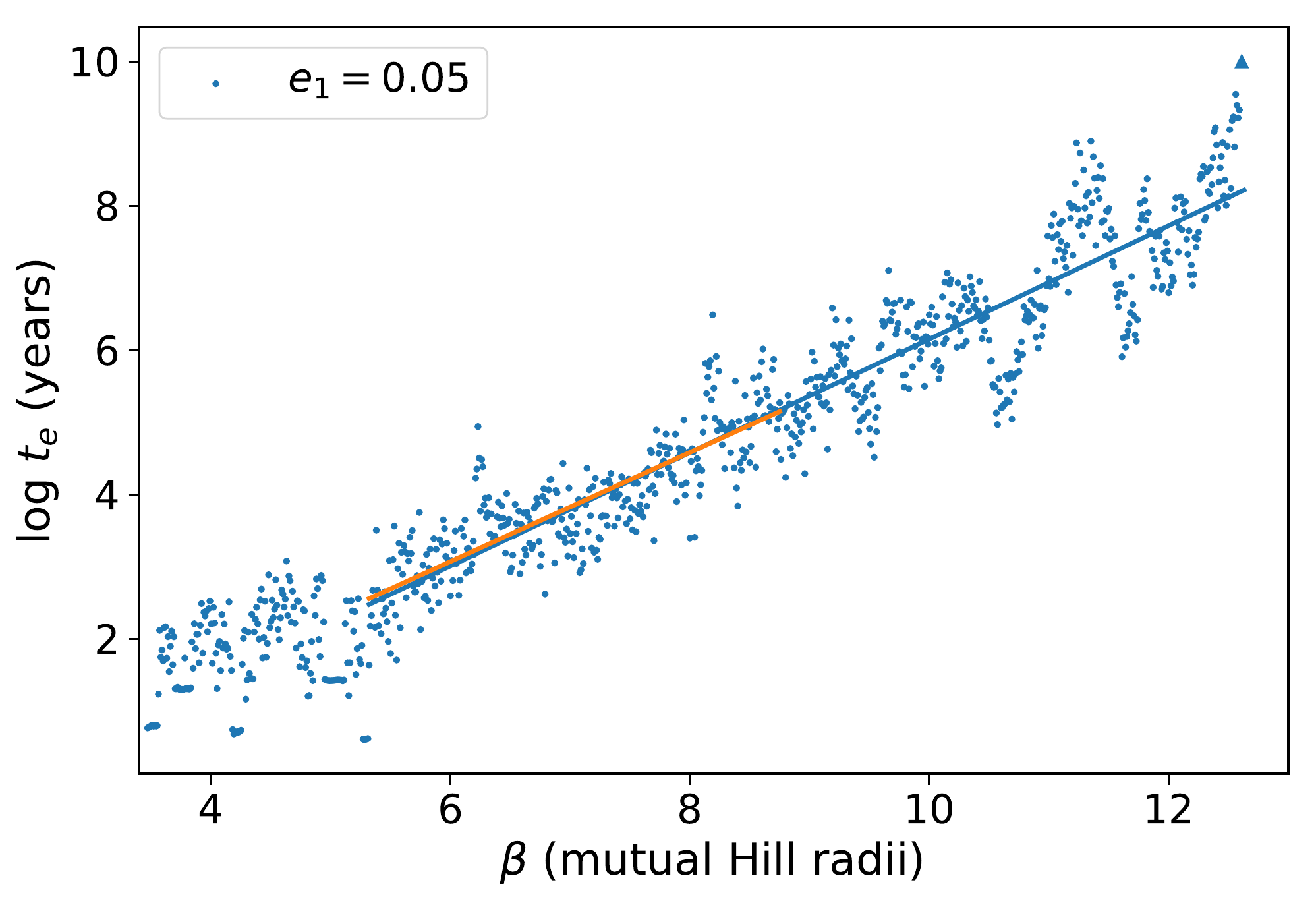} 
\caption{Close encounter times for systems whose innermost planet starts out with eccentricity $e_1=0.05$. In blue the corresponding regression line, starting just after the widest separation system that has a collision during the first synodic period; in orange, the regression line for the same interval in $\beta$ as the orange line in Figure \ref{fig:e001_reg}}
\label{fig:e005_reg}  
\end{center}  
\end{figure}

\begin{figure}[H]
\begin{center}  
\includegraphics[width=1.0\textwidth]{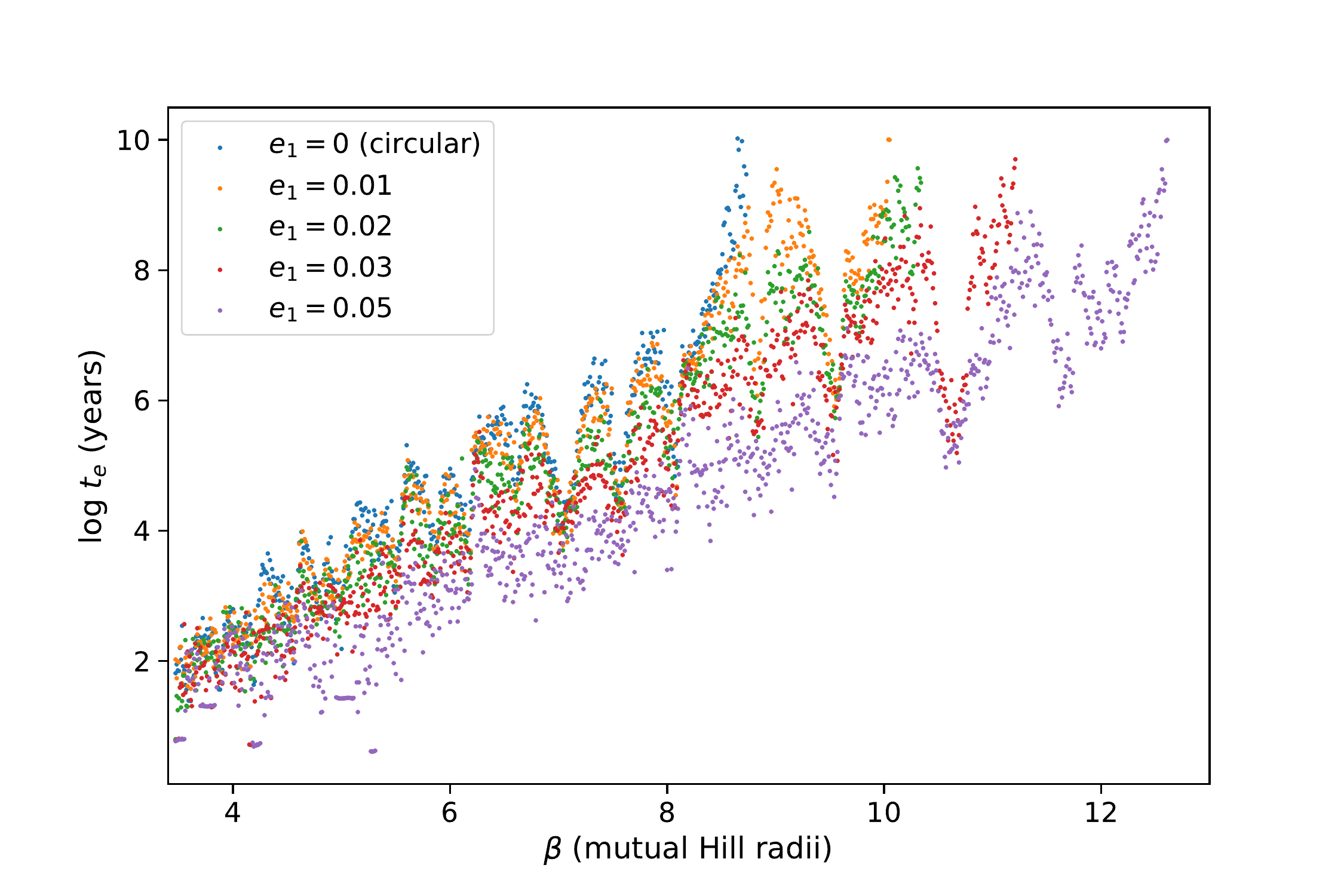} 
\caption{Close encounter times for initial $e_1=0, 0.01, 0.02, 0.03$, and $0.05$, respectively.}
\label{fig:all_eccs_e1}  
\end{center}  
\end{figure}

As the eccentricity increases, we expect the lifetimes to decrease for fixed $\beta$. This is indeed what happens, and Figure \ref{fig:e005_reg} shows the behavior of the highest eccentricity inner planet systems we simulated, $e_1=0.05$.
The persistence of a log-linear relationship between initial separation and close encounter times for eccentricities at least up to $0.05$ can be seen in Figure \ref{fig:all_eccs_e1}, with various slopes. Several systems (one for $e_1=0.02$, eight for $e_1=0.03$, and twenty-two for $e_1=0.05$), suffered a close approach extremely rapidly, with lifetimes less than ten years. Those encounters are between planets $1$ and $2$ and occur during their first synodic period: given our choices of initial longitudes and periapse angle of the inner planet, the first passage occurs when the inner planet is near apoapse. Those excessively short lifetimes are thus direct consequences of our initial longitudes.



\FloatBarrier

We conclude that for any given value of $e_1$ simulated, system lifetime tends to roughly follow the log-linear relationship with $\beta$ of the form given in Eq.~\ref{eq:power law}, although the coefficients differ from those of initially circular orbits. 

We now quantify the differences in close encounter times by computing their average over a range of three Hill radii, ending with the last point available for the circular case (i.e., the fifth initially circular system exceeding two billion years of stability), which amounts to the range $\beta\in[5.74, 8.73]$. 

\begin{figure}[!htb]
\begin{center}  

\includegraphics[width=1.0\textwidth]{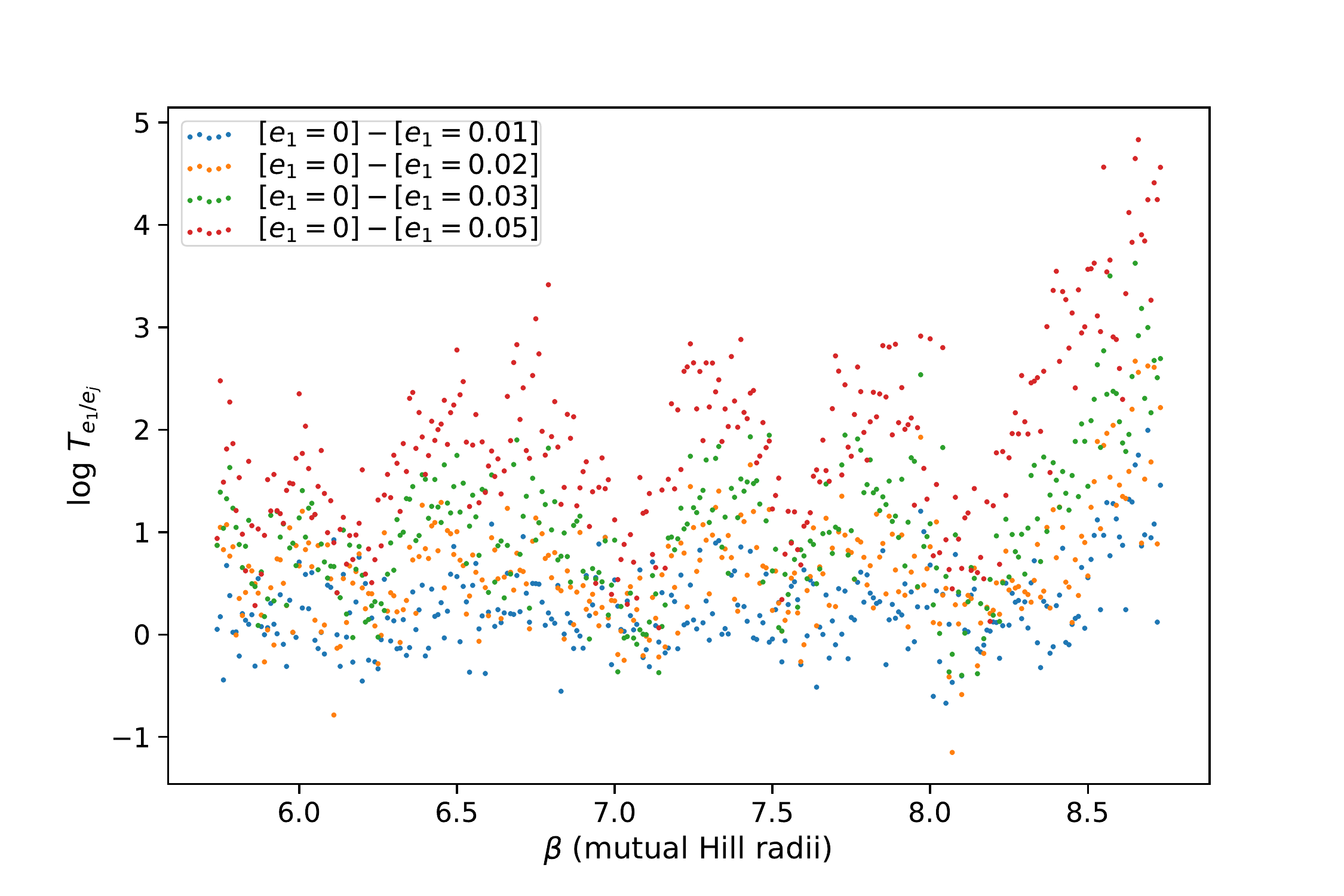}
\caption{Close encounter time differences between simulations with all planets starting on circular orbits from those with $e_1=0.01, e_1=0.02$, and $e_1=0.05$. In the legend, $[e_1=0]$ is a time unit and means $\log t_e[e_1=0]$; and so on. The average time difference between the circular system and the corresponding (same $\beta$) eccentric system is given in Table \ref{tab:multiplicative_factors_e_1_last300}.}
\label{fig:differences_planet_1} 
\end{center}

\end{figure}

Thus in Figure \ref{fig:differences_planet_1}, each point represents the log-difference in close encounter time between one of the four eccentricity cases studied and the initially circular system. Not surprisingly, the higher the eccentricity, the larger the close encounter time differences. The four averages over the whole range are recorded in Table \ref{tab:multiplicative_factors_e_1_last300}. 

We omit similar plots for other planets eccentric -- the ratios follow the same pattern for comparisons of different eccentricities of the same planet. However, an interesting pattern is found for the comparison of the average close encounter time between \textit{different} planets starting out with the \textit{same} eccentricity (Figure \ref{fig:differences_e005}).

\begin{table}[!htb]
\begin{center}
    \begin{tabular}{ | c | c | c | c | c | c | c |}
    \hline
    circular & $e_1=0.01$ &  $e_1=0.02$ & $e_1=0.03$ & $e_1=0.05$ & $e_{\rm align}=0.05$ \\ \hline \hline
    $1$ & $0.543$ & 0.237 & 0.091 & 0.013 & 0.331    \\ \hline
    
    \end{tabular}
\caption{The ``average multiplicative factor" in close encounter time difference for each batch of systems compared to the circular systems.  We compute this factor by exponentiating the average of the differences in the logarithms of the encounter time of simulations with the same value of $\beta$, within a specified range in $\beta$. The range consists of the last $300$ overlapping points $\beta\in [5.74, 8.73]$.  The batch of simulations labeled $e_{\rm align}=0.05$ has all planets on eccentric and aligned orbits; see Section \ref{sec: all planets eccentric and aligned orbits} for details. } 
\label{tab:multiplicative_factors_e_1_last300}
\end{center}
\end{table}

\section{Outermost planet eccentric} \label{sec: outer planet eccentric}
In this section, we present simulations analogous to those in Section \ref{sec: innermost planet eccentric} except that the initial eccentricity is given to the outer planet.  We expect results qualitatively similar to those in which the inner planet is eccentric, but quantitative differences may exist. 


We introduce here the Angular Momentum Deficit (AMD).  
For the planar problem that we are studying herein, the AMD of a planet is defined as the difference between the angular momentum it would have on a circular orbit of radius equal to its semi-major axis and its actual orbital angular momentum:
\begin{equation}
\text{AMD} = m\sqrt{GMa}(1-\sqrt{1-e^2}).
\end{equation}

\noindent The total AMD of a system is not changed by secular interactions among the planets, and thus the close encounter times of our systems may be strongly  correlated with their  initial AMD.  (See also \cite{laskar1997large}, and \cite{laskar2017amd} for an analytically derived AMD-stability criterion.) Comparing the AMD of a system with inner planet eccentric with one having planets at the same semi-major axes but the outer planet having the same eccentricity, we see that because the outer planet has larger $a$, the system with the outer planet eccentric has larger AMD.

What role does angular momentum deficit play in the destabilization of the closely-packed planetary systems that we are investigating? We look to answer this question in the following way. By giving nonzero eccentricity not to the innermost planet, but to a different one instead, we can calculate the expected outcome if AMD is at play. In particular, if the initial value of the AMD is key, then there should be no systematic difference in lifetimes if the nonzero initial eccentricity of the new planet is adjusted relative to its respective initial semi-major axis such that it matches the initial AMD of the innermost planet, whereas systems at the same $\beta$ with the outermost planet eccentric would on average be shorter-lived than those with the innermost planet having the same value of eccentricity. 


We performed six batches of simulations with only the outer planet starting on an eccentric orbit.  Three of these used fixed values of $e_5$. The other three adjusted $e_5$ to  match the AMD of the corresponding system where the innermost planet starts out with nonzero initial eccentricity $e_1$.
 For each matched AMD simulation (i.e., each value of $\beta$), the initial eccentricity, $e_5$, is adjusted such that the initial angular momentum deficit of that system is the same as that for the $e_1\neq 0$ system:
\begin{equation}
m_1\sqrt{GM_\odot a_{1}}\Bigl(1-\sqrt{1-e_{1}^2}\Bigr) = m_5\sqrt{GM_\odot a_{5}}\Bigl(1-\sqrt{1-e_{5}^2}\Bigr). 
\end{equation}

This equation can be solved for $e_5$. Taking into account equal planetary masses $m_1 = m_5$, we have

\begin{equation} \label{eq: e5amditoe1}
e_5 = \Bigl[1 - \Bigl(1 - \sqrt{\frac{a_1}{a_5}}\Bigl(1 - \sqrt{1 - e_1^2}\Bigr)\Bigr)^2\Bigr]^{1/2}.
\end{equation}

\noindent We use Equation (\ref{eq: separation2}) to get the semi-major axes ratio, and for simplicity define
\begin{equation}
X\equiv\frac{1}{2}\Bigl(\frac{2M_{\oplus}}{3M_{\odot}}\Bigr)^{1/3} \approx 0.0063.
\end{equation}
We recursively apply Equation (\ref{eq: separation2}) giving us $a_5$ in terms of $a_1$:
\begin{equation}
a_5 = \Bigl(\frac{1+\beta X}{1-\beta X}\Bigr)^4 a_1.
\end{equation}

\noindent Plugging this back into Equation (\ref{eq: e5amditoe1}), we finally have
\begin{equation}
e_5 = \Bigl[1 - \Bigl(1 - \Bigl(\frac{1+\beta X}{1-\beta X}\Bigr)^{-2}\Bigl(1 - \sqrt{1 - e_1^2}\Bigr)\Bigr)^2\Bigr]^{1/2}.
\end{equation}

\noindent The choice of initial $e_5$ thus depends on both the initial separation $\beta$ and the corresponding system's initial eccentricity $e_1$ of the innermost planet.

\begin{figure}[!htb]
\begin{center}  
\includegraphics[width=1.00\textwidth]{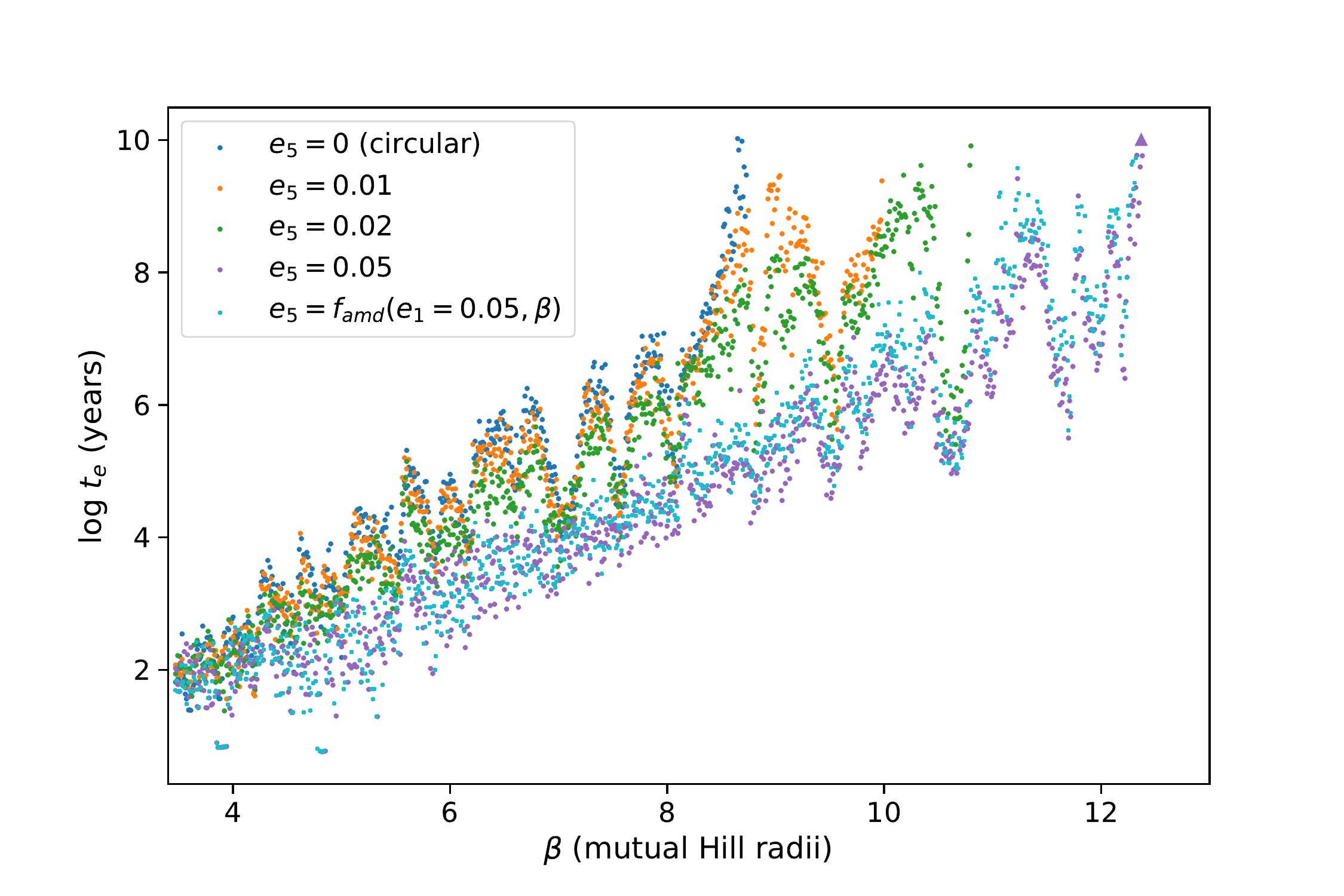} 
\caption{Close encounter times for initial $e_5=0, 0.01, 0.02, 0.05$, and the AMD-adjusted $e_5$. The AMD-adjusted systems tend to be slightly shorter-lived than those starting out at $e_5=0.05$ for all $\beta$, which is to be expected, since the adjustment puts the initial $e_5$ eccentricity on an interval going from $0.048$ for $\beta=3.47$ down to $0.043$ for $\beta = 12.32$, thus always remaining smaller than but close to $0.05$ eccentricity. However, the interesting part is to compare those systems with those whose innermost planet begins at $e_1=0.05$, since their initial AMD is identical by construction, while the initial eccentricity is in different planets. This figure is the counterpart to Figure \ref{fig:all_eccs_e1}, where we plotted the lifetimes of runs with different eccentricities for the innermost planet.}
\label{fig:all_eccs_e5}  
\end{center}  
\end{figure}
\FloatBarrier

Table \ref{tab:all_fits} presents coefficients of the exponential fits to all six batches of systems with just the outer planet eccentric. Figure \ref{fig:all_eccs_e5} shows the close encounter times, again as a function of $\beta$, of all three batches with constant values of the outermost planet's initial eccentricity, $e_5$, as well as the batch with $e_5$ chosen to match the AMD of the run with $e_1=0.05$.  General trends are, as expected, similar to those seen for analogous batches with the inner planet eccentric.

\begin{figure}[!htb]
\begin{center}  
\includegraphics[width=1.0\textwidth]{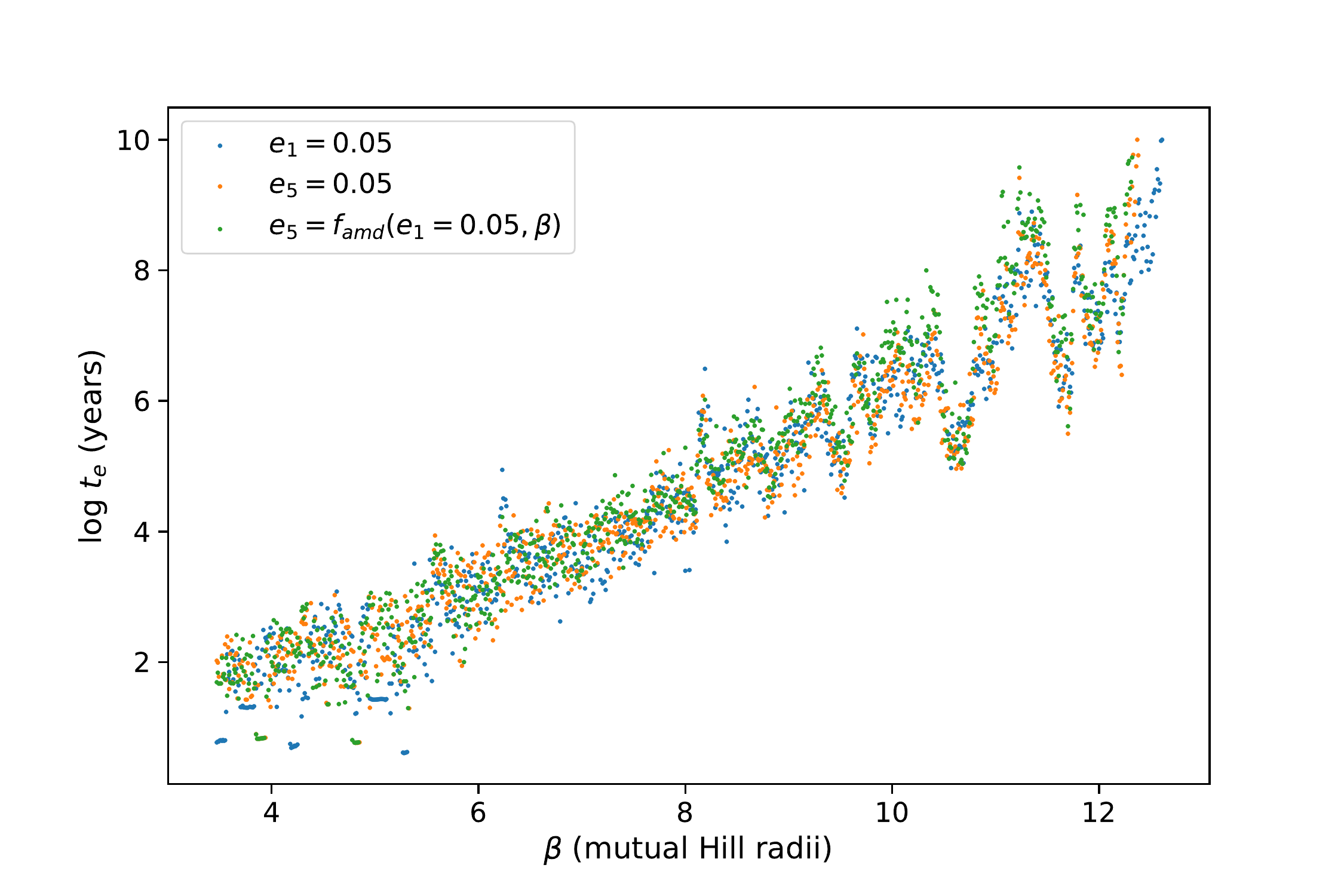} 
\caption{Close encounter times of systems with initial $e_1=0.05, e_5=0.05$, and $e_5=f_{\rm amd}(e_1=0.05, \beta)$.}
\label{fig:eccs_005_amd}  
\end{center}  
\end{figure}
\FloatBarrier

Figure \ref{fig:eccs_005_amd} compares lifetimes for batches of systems with either the inner or outer planet beginning with eccentricity $e\approx0.05$, and Figure \ref{fig:differences_e005} shows difference in close encounter time between innermost and outermost planet for both the standard $e_5$ batch and the AMD adjusted batch, complementing the comparison of different eccentricities for the same planet shown in Figure \ref{fig:differences_planet_1}. There, we focused on the change in close encounter times for different initial eccentricities of the same (innermost) planet, whereas here our focus is on the change in close encounter times for different planets with the same initial eccentricity (we choose the largest eccentricities, $e\approx0.05$).

Figure \ref{fig:differences_e005} shows that the systems with $e_5\approx0.05$ and those with $e_1=0.05$ have similar lifetimes. Only the 300 most widely-spaced systems for which we have data, $\beta \in [9.33, 12.32]$ are shown here, because the long-lived systems with these separations are more relevant to observed exoplanet systems than the more closely-spaced systems that we modeled. The  $e_5=0.05$ systems on average have a somewhat shorter close encounter time than those with $e_1=0.05$, but the difference is small, with the average multiplicative factor being 0.881. On the other hand, the red points for the AMD-adjusted batch lie significantly below the line of zero difference: The AMD-adjusted systems live much longer than the $e_1=0.05$ systems, with an average ratio of 2.393. This is mostly due to the smaller initial eccentricities of those systems: they range from $e_5=0.0444$ at $\beta=9.33$ to $e_5=0.0428$ at $\beta=12.32$.

\FloatBarrier

The slopes, $b'$, and intercepts, $c'$, of exponential fits to all batches of  runs are given in Table \ref{tab:all_fits}, using a separation shifted by the critical separation for Hill stability of two-planet systems (eq.~\ref{eq:beta prime}. Table \ref{tab:all_fits} also gives the values of
  $\sigma_{b'}$ and  $\sigma_{c'}$, the standard errors of our estimates of the slopes and intercepts). For most batches, we considered multiple intervals in $\beta$. We begin by comparing our numbers with circular initial orbits (lines one and two). We then record the data for all batches, starting from the first point beyond which there are no more invalid points, until the fifth system within the batch reaching two billion years of lifetime. Next, we compare innermost planet eccentricities including $e_1=0.05$, before comparing all batches \textit{except} any single $e_i=0.05$ (but including $e_{\rm align} =0.05)$. Next are the batches with one planet having eccentricity $\approx 0.05$. Finally, fit coefficients to batches of systems with all planets eccentric and randomized periapses within a given range are given.

\newgeometry{left=10mm, right=10mm, top=0.1cm,bottom=0.1cm}
\begin{table}
    \begin{tabular}{ | l | c | c | c | c | c |}
    \hline
    Eccentricity & Range &   \textit{b}$'$  & $\sigma_{b'}$ & \textit{c}$'$ & $\sigma_{c'}$\\ \hline \hline
    $0$ & $3.47 \leq \beta \le 8.73$ & 1.080 & 0.019 & 1.85 & 0.06    \\ \hline
    $e_1=0.01$ & $3.47 \leq \beta \le 8.73$ & 0.999 & 0.016 & 1.87 & 0.05  \\ \hline 
     & last invalid $\to$ fifth beyond $2\times 10^9$ \\ \hline
    $e_1=0.01$ & $3.47 \leq \beta \le 10.05$ & 1.030 & 0.014 & 1.82 & 0.05   \\ \hline
    $e_1=0.02$ & $3.48 \leq \beta \le 10.34$ & 0.984 & 0.010 & 1.61 & 0.04 \\ \hline
    $e_1=0.03$ & $4.19 \leq \beta \le 11.21$  & 0.867 & 0.011 & 1.58 & 0.05 \\ \hline
    $e_1=0.05$ & $5.32 \leq \beta \le 12.61$ & $0.785$  & 0.010  & $1.02$ & 0.06  \\ \hline
    $e_2=0.05$ & $5.49 \leq \beta \le 12.99$ & $0.827$ & 0.010  & $0.52$ & 0.06 \\ \hline
    $e_3=0.01$ & $3.47 \leq \beta \le 10.09$ & 1.017  & 0.012  & $1.82$ & 0.05 \\ \hline
    $e_3=0.02$ & $4.00 \leq \beta \le 11.09$ & 0.921 & 0.012  & $1.64$ & 0.06\\ \hline
    $e_3=0.05$ & $5.07 \leq \beta \le 12.99$ & 0.810 & 0.009  & $0.56$ & 0.06 \\ \hline
    $e_5=0.01$ & $3.47 \leq \beta \le 9.98$ & 1.016  & 0.014& $1.84$ & 0.05 \\ \hline
    $e_5=0.02$ & $3.47 \leq \beta \le 10.80$ & 0.911 & 0.012 &  $1.79$ & 0.05 \\ \hline
    $e_5=0.05$ & $4.86 \leq \beta \le 12.38$ & 0.759 & 0.010  & $1.13$ & 0.06\\ \hline
    $e_5=e_1^{\rm amd}|_{0.01}$ & $3.47 \leq \beta \le 10.05$ & 0.981 & 0.013  & $1.93$ & 0.05\\ \hline
    $e_5=e_1^{\rm amd}|_{0.02}$ & $3.47 \leq \beta \le 10.30$ & 0.983 & 0.011  & $1.68$ & 0.05  \\ \hline
    $e_5=e_1^{\rm amd}|_{0.05}$ & $4.85 \leq \beta \le 12.32$ & 0.826 & 0.011   & $1.02$ & 0.06  \\ \hline
    $e_{\rm align}=0.05$ & $3.47 \leq \beta \le 10.08$ & 0.837 & 0.016  & $2.18$ & 0.06 \\ \hline
    $e_1\neq 0$: & overlapping range: \\ \hline
    $e_1=0.01$ & $5.32 \leq \beta \le 10.05$ & 1.056 & 0.025 & 1.69 & 0.11  \\ \hline 
    $e_1=0.02$ & $5.32 \leq \beta \le 10.05$ & 0.973 & 0.020 & 1.60 & 0.09 \\ \hline
    $e_1=0.03$ & $5.32 \leq \beta \le 10.05$  & 0.907 & 0.017 & 1.44 & 0.08 \\ \hline
    $e_1=0.05$ & $5.32 \leq \beta \le 10.05$ & $0.767$  & 0.016  & $1.11$ & 0.07  \\ \hline
    
    all except $e\approx0.05$: & overlapping range:\\ \hline
    $e_1=0.01$ & $4.21 \leq \beta \le 10.05$ & 1.033 & 0.017 & 1.80 & 0.07  \\ \hline
    $e_1=0.02$ & $4.21 \leq \beta \le 10.05$ & 0.969 & 0.014 & 1.63 & 0.05 \\ \hline
    $e_1=0.03$ & $4.21 \leq \beta \le 10.05$  & 0.893 & 0.012 & 1.50 & 0.05 \\ \hline
    $e_3=0.01$ & $4.21 \leq \beta \le 10.05$ & 1.005  & 0.016  & $1.87$ & 0.06 \\ \hline
    $e_3=0.02$ & $4.21 \leq \beta \le 10.05$ & 0.954 & 0.013  & $1.56$ & 0.05\\ \hline
    $e_5=0.01$ & $4.21 \leq \beta \le 10.05$ & 1.009  & 0.017& $1.88$ & 0.07 \\ \hline
    $e_5=0.02$ & $4.21 \leq \beta \le 10.05$ & 0.951 & 0.013 &  $1.69$ & 0.05 \\ \hline
    $e_5=e_1^{\rm amd}|_{0.01}$ & $4.21 \leq \beta \le 10.05$ & 0.963 & 0.017  & $2.02$ & 0.07\\ \hline
    $e_5=e_1^{\rm amd}|_{0.02}$ & $4.21 \leq \beta \le 10.05$ & 0.961 & 0.015  & $1.72$ & 0.06  \\ \hline
    $e_{\rm align}=0.05$ & $4.21 \leq \beta \le 10.05$ & 0.790 & 0.020  & $2.37$ & 0.08 \\ \hline
    
    $e\approx0.05$: & overlapping range:\\ \hline
    $e_1=0.05$ & $5.49 \leq \beta \le 12.32$ & $0.755$  & 0.011  & $1.16$ & 0.06  \\ \hline
    $e_2=0.05$ & $5.49 \leq \beta \le 12.32$ & $0.778$ & 0.011  & $0.73$ & 0.06 \\ \hline
    $e_3=0.05$ & $5.49 \leq \beta \le 12.32$ & 0.765 & 0.011  & $0.76$ & 0.07 \\ \hline
    $e_5=0.05$ & $5.49 \leq \beta \le 12.32$ & 0.742 & 0.011  & $1.22$ & 0.07\\ \hline
    $e_5=e_1^{\rm amd}|_{0.05}$ & $5.49 \leq \beta \le 12.32$ & 0.829 & 0.012 & $1.00$ & 0.07  \\ \hline
    all $e\approx0.05$, randomized periapses: & range:\\ \hline
    $e_{all}|_{\ang{45}}=0.05$ & $3.47 \leq \beta \le 11.24$ & 0.749 & 0.012 & $1.81$ & 0.06  \\ \hline
    $e_{all}|_{\ang{90}}=0.05$ & $3.47 \leq \beta \le 12.60$ & 0.628 & 0.009 & $1.48$ & 0.05  \\ \hline
    $e_{all}|_{\ang{135}}=0.05$ & $3.47 \leq \beta \le 13.10$ & 0.593 & 0.009 & $1.04$ & 0.05  \\ \hline
    $e_{all}|_{\ang{180}}=0.05$ & $3.47 \leq \beta \le 14.19$ & 0.532 & 0.008 & $0.95$ & 0.05  \\ \hline
    $e_{all}|_{\ang{360}}=0.05$ & $3.47 \leq \beta \le 15.18$ & 0.479 & 0.006 & $0.69$ & 0.04  \\ \hline
    
\end{tabular}
\caption{Slopes and intercepts and their standard deviations for log-linear fits to lifetimes of simulated batches of systems.}  
\label{tab:all_fits} 
\end{table}

\restoregeometry

\FloatBarrier

\section{One intermediate planet eccentric} \label{sec: one intermediate planet eccentric}
\begin{figure}[!htb]
\begin{center}  
\includegraphics[width=0.99\textwidth]{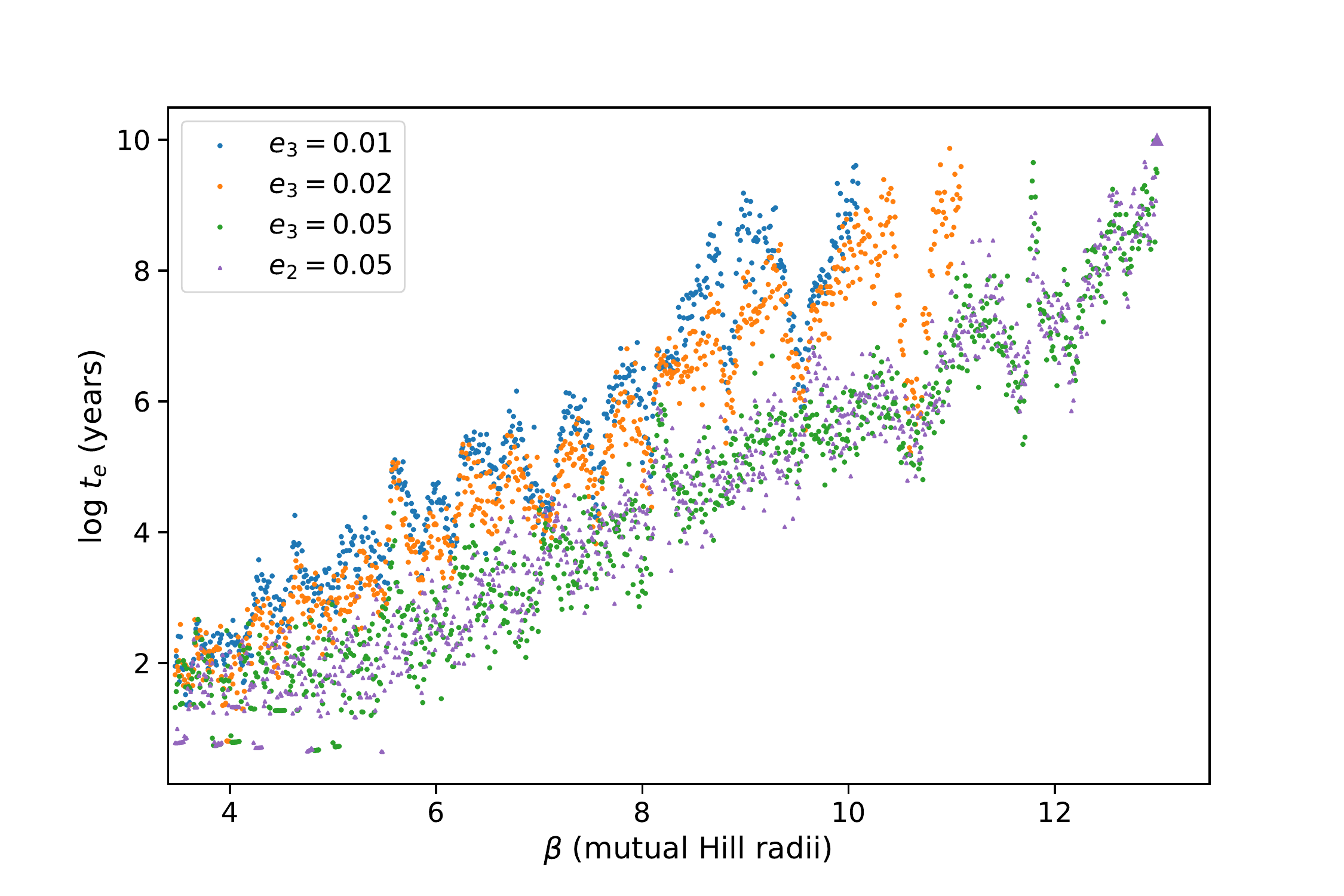} 
\caption{Close encounter times for $e_3=0.01, e_3=0.02,$ and $e_3=0.05$, and, in purple, $e_2=0.05$. }
\label{fig:superposition_e3001_e3002_e3005_e2005}  
\end{center}  
\end{figure}

We close the main part of this investigation by considering systems in which one of the intermediate planets is initially eccentric. For most of these runs we choose the middle (third) planet initially eccentric, and also included a run with $e_2=0.05$. Figure \ref{fig:superposition_e3001_e3002_e3005_e2005} confirms the log-linear relationship for those planetary systems. 

\begin{figure}[!htb]
\begin{center}  
\includegraphics[width=1.0\textwidth]{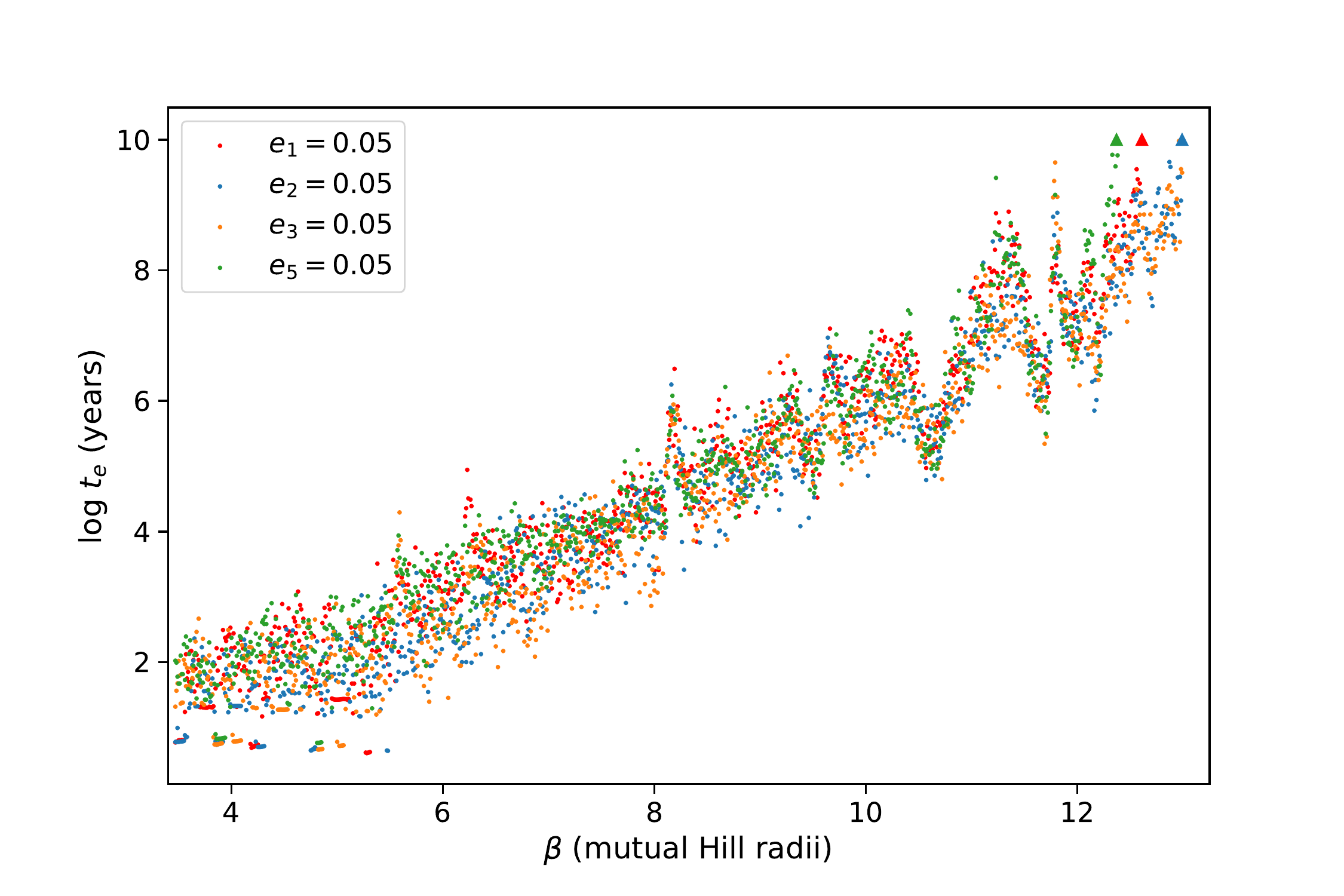} 
\caption{Close encounter times for $e_1=0.05, e_2=0.05, e_3=0.05$, and $e_5=0.05$.}
\label{fig:e1e3e5005}  
\end{center}  
\end{figure}

\begin{figure}[!htb]
\begin{center}  
\includegraphics[width=1.0\textwidth]{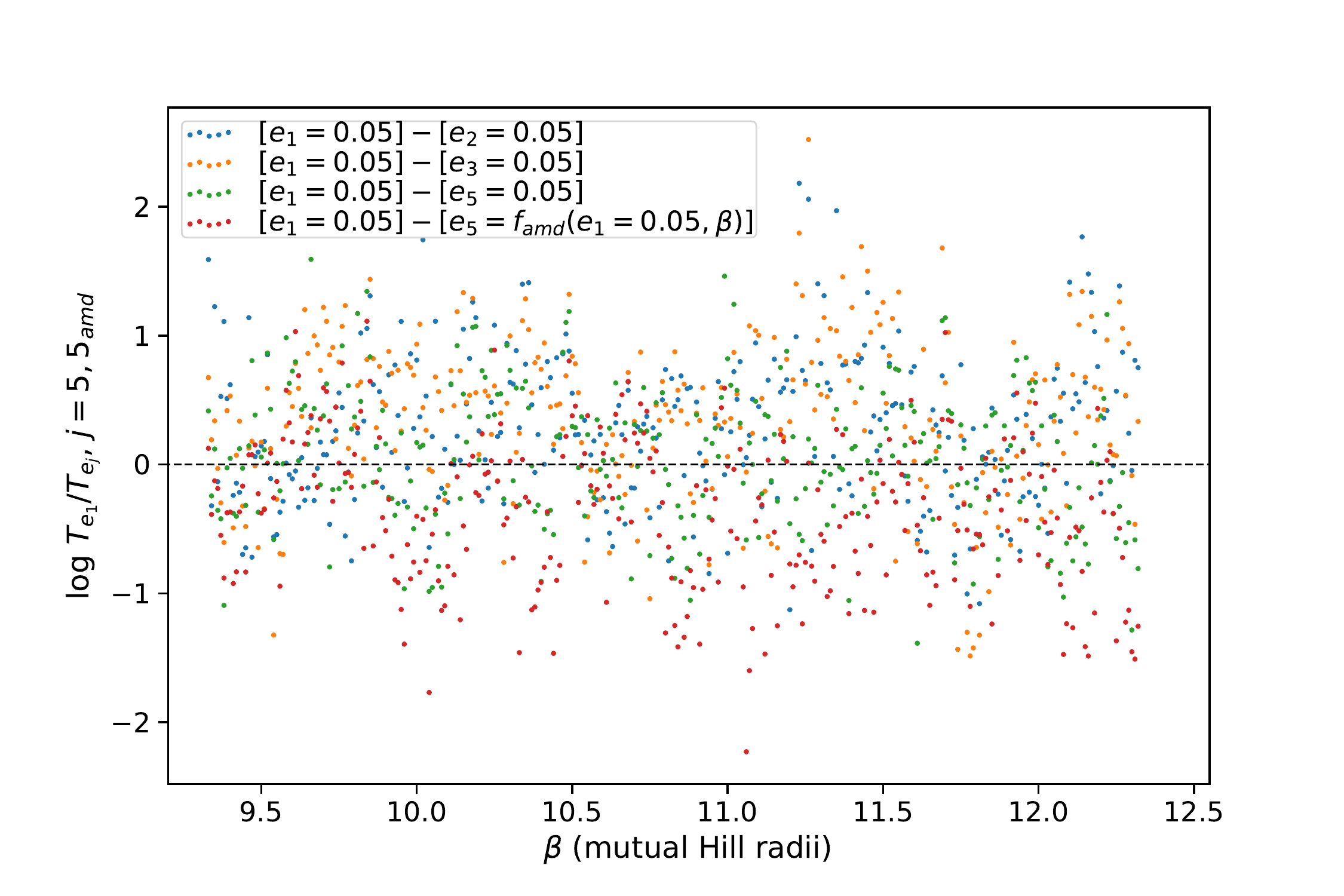} 
\caption{Close encounter differences between simulations with initial $e_1=0.05$ and those with $e_2=0.05$, $e_3=0.05$, $e_5=0.05$, or $e_5=f_{\rm amd}(e_1=0.05, \beta)$. This plot represents the quantitative comparison between the systems shown in Figures \ref{fig:eccs_005_amd} and \ref{fig:e1e3e5005}. Here, the average multiplicative factors in close encounter time compared to the $e_1=0.05$ system are, in the plotting order, 0.525, 0.436, 0.881, and 2.393, respectively. Thus, the systems with intermediate planet eccentric (both $e_2$ and $e_3$) are about half as long-lived on average as those without two planetary neighbors.}
\label{fig:differences_e005}  
\end{center}  
\end{figure}

We  compare the $0.05$ systems in Figure \ref{fig:e1e3e5005}. We find slightly lower close encounter times for eccentric intermediate planets, as shown in Figure \ref{fig:differences_e005}. We thus confirm that a log-linear relationship continues to hold for intermediate planets eccentric, and find the close encounter times decrease on average by roughly a factor of two when the initially eccentric planet has both interior and exterior neighbors. The slopes ($b'$) and intercepts ($c'$) fit for the log-linear relationship in the overlap range $\beta \in[5.49, 12.32]$ are similar for all four runs, with the systems with the initially eccentric planet in the middle having slightly steeper slopes but somewhat smaller intercept values than those with the inner or outer planet eccentric.  The $e_5 = e_1^{amd}|_{0.05}$ batch has a steeper slope because the eccentricity drops (albeit slowly) as $\beta$ is increased.

\FloatBarrier

\section{All planets on eccentric orbits} \label{sec: all planets eccentric and aligned orbits}
In the previous sections, we found that for eccentricities up to $0.05$, systems with one planet eccentric follow a log-linear behavior as a function of initial separation, and this is true no matter which planet starts out with the nonzero eccentricity. Here, we consider systems with all planets on initially eccentric orbits with $e=0.05$. In one batch, all periapses angles are initially aligned, $e_{\rm align}=0.05$, whereas the others have periapse angles that are randomly-drawn over specified ranges in azimuth of width $45^\circ-360^\circ$. 

\begin{figure}[!htb]
\begin{center}  
\includegraphics[width=1.0\textwidth]{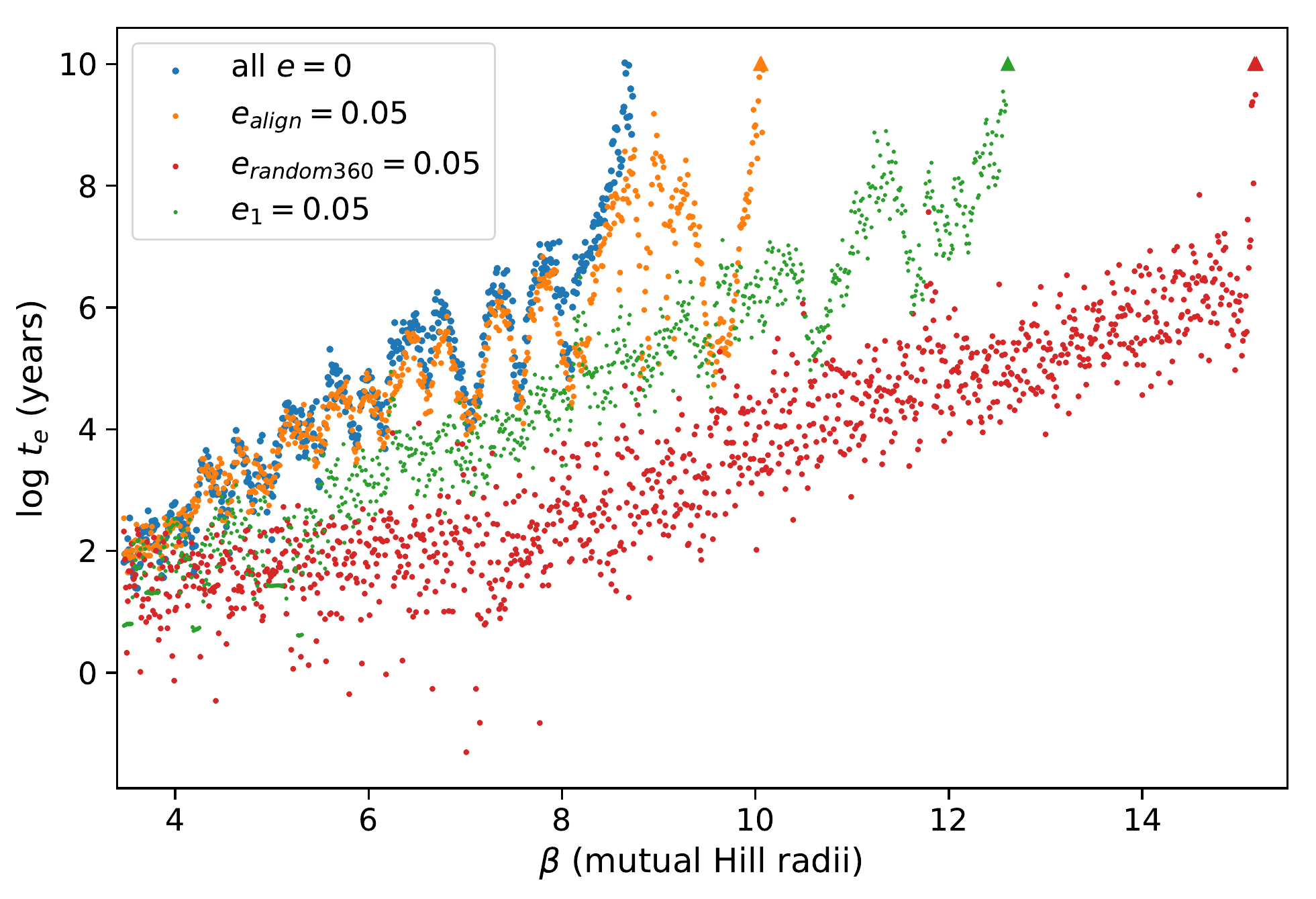} 
\caption{Superposition of close encounter times for the circular systems, all planets eccentric and aligned, all planets eccentric with randomized initial periapse angles, and the $e_1=0.05$ systems. }
\label{fig:eall005_e0_e_align_e1005}  
\end{center}  
\end{figure}

Figure \ref{fig:eall005_e0_e_align_e1005} compares our results with previously-shown results for circular orbits and for $e_1 =0.05$. 
The difference between the systems with just one planet eccentric and all of them eccentric is stark: when all planets start out aligned and at the same eccentricity of $e_{\rm align}=0.05$, the system lifetimes are considerably longer than those with just one planet having $e=0.05$, reaching the stopping point of five runs lasting longer than $2 \times10^9$ years at separation similar to the batches with one planet having initial $e=0.01$. In contrast, randomized initial angular parameters dramatically reduce the close encounter times, and no resonance-related structure is appearing up until at least $\beta = 15$.

It is clear that the initial difference between the eccentricity vectors of the planets (a quantity sometimes referred to as relative eccentricity of the planets) is more important than individual planetary eccentricity values. This result is in qualitative agreement with \cite{0004-637X-807-1-44}'s finding that the minimum initial approach distance of orbits in a system of mildly eccentric planets is a key factor in determining the system's stability. However, as the lifetimes of the systems with aligned eccentric orbits are somewhat less than those of initially circular orbits for the largest orbital separations that we investigated, we find that for a given initial minimum separation between orbits, larger eccentricities typically reduce system lifetimes by a smallish factor.
Also, the lifetimes of aligned eccentric runs are not as well-approximated by a log-linear fit as other configurations that we studied. Comparison with two other sets that reached five long-lived runs at similar values of $\beta$ (to minimize effects of including differing resonances at different ranges), ($e_1=0.01$ and $e_3=0.01$) shows a smaller coefficient of determination, $R^2$, compared to the latter two batches, $0.803$ vs. $0.898$ and $0.910$, respectively.


\FloatBarrier

\begin{figure}[!htb]
\begin{center}  
\includegraphics[width=1.0\textwidth]{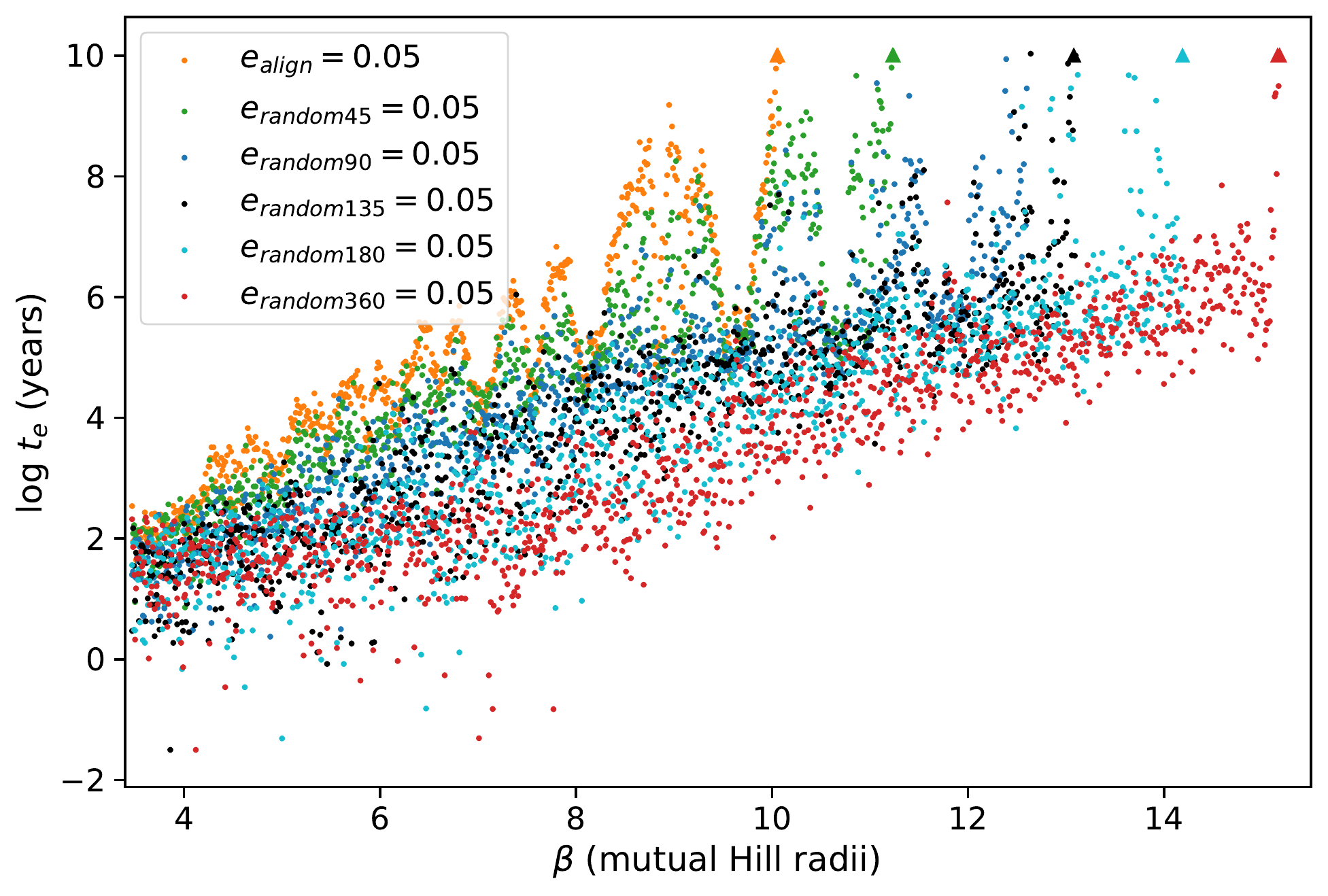} 
\caption{Superposition of close encounter times for systems with periapses aligned and those with periapses randomized within $45^\circ, 90^\circ, 135^\circ, 180^\circ$ and $360^\circ$.}
\label{fig:eall005angles}  
\end{center}  
\end{figure}

The substantial differences in lifetimes of systems with all planets having initial $e = 0.05$ for aligned and randomly-oriented initial longitudes of periapse seen in Fig.~\ref{fig:eall005_e0_e_align_e1005} motivated us to investigate intermediate cases in which initial periapse angles are spread over a limited range in angle. Results are presented Figure \ref{fig:eall005angles}. We see a gradual transition between the aligned and fully random results for intermediate amounts of periapse dispersion. Nonetheless, all batches approximately follow an exponential relationship between initial spacing and encounter time.

\FloatBarrier

\section{Exponential Fitting Coefficients for Differing Batches of Run}\label{sec:fits}
\begin{figure}[!htb]
\begin{center}  
\includegraphics[width=1.0\textwidth]{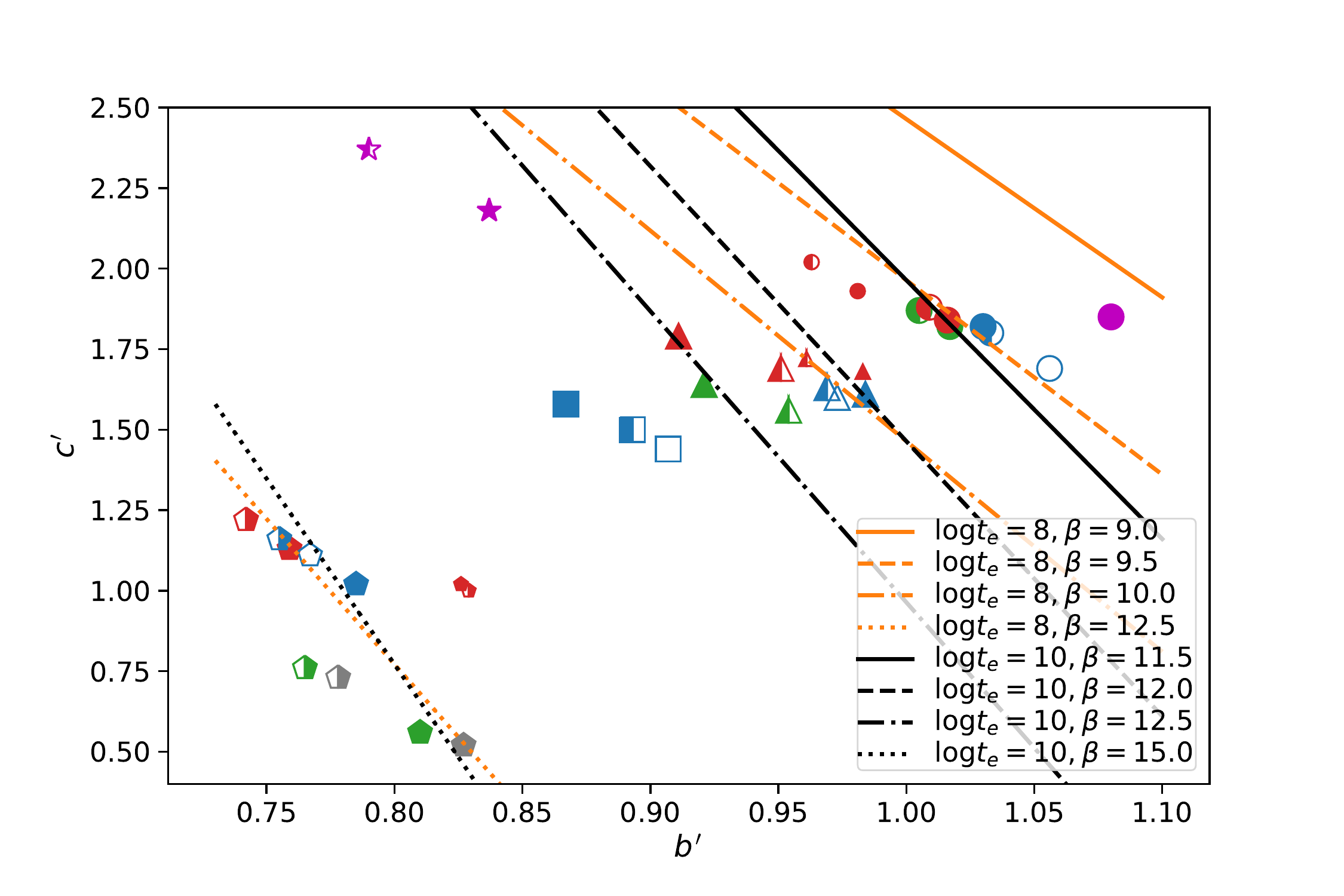} 
\caption{Slopes and intercepts for the exponential fits to the lifetimes of systems in various batches of runs over the ranges in $\beta$ specified below. The shape of the symbol shows the initial eccentricity of the specified planet(s), with circles representing  both the circular runs and those with $e = 0.01$, upward-pointing triangles representing $e=0.02$, squares $e=0.03$ and pentagons $e=0.05$. We used a star shape for the all planets eccentric ($e=0.05$) and aligned systems, since the relevant feature for their lifetimes are not the individual planets' eccentricities, but the alignment of their periapses. Blue symbols denote batches with the innermost planet initially eccentric, grey (green) symbols denote those with the second (third) initially eccentric, red symbols signify that the outermost planet is initially eccentric, and magenta symbols have either all planets circular or all planets eccentric and aligned. Filled symbols correspond to fits over the full range from just after the widest separation system that has a collision during the first synodic period (or $\beta=3.47$ if no system in the batch suffered this fate) through the fifth system to survive for $>2\times10^9$ years. 
Open and semi-open symbols represent restricted ranges, with empty symbols fit over the range $5.32 \leq \beta \le 10.05$, left-filled symbols relying on data from the range $4.21 \leq \beta \le 10.05$ and right-filled symbols for the range is $5.49 \leq \beta \le 12.32$.}
 The curves represent contours of constant $\beta$ for log-linear fit lifetimes of 10$^8$ and 10$^{10}$ years.

\label{fig:coeffs}  
\end{center}  
\end{figure}

Figure \ref{fig:coeffs} plots most of the slopes and intercepts from Table \ref{tab:all_fits} in the $(b',c')$ plane. We also show lines of constant lifetimes ($10^8$ years and  $10^{10}$ years) for specified values of $\beta$ to facilitate comparison of the stability of differing batches. Details are described in the caption. The systems with initial eccentricity of $0.01$ (circular symbols, apart from one that is magenta) lie near the line corresponding to lifetimes of $10^8$ years for $\beta=9.5$. Similarly, the most eccentric systems in our simulations (one planet having initial $e\approx0.05$, pentagons) accumulate around the lines of highest drawn separation, both at $10^8$ years ($\beta=12.5$) and $10^{10}$ years ($\beta=12.5$). Generally, points in the lower left correspond to batches of runs that are relatively short-lived, whereas those in the upper right correspond to long-lived batches. 
Results for different ranges of the same set (filled, open and partly filled points of the same size, shape and color) don't differ by much, and the differences that are present don't trend in a systematic manner with with the size of the fit region.  Slopes are less steep for larger $e$ because there is not much difference in $t_c$ for closely-spaced orbits but substantial differences in lifetime are present for large values of $\beta$.  There is only a slight reduction in $c'$ (the fit lifetime of the most closely-spaced systems) as initial eccentricity is increased from 0 to 0.03,  but it decreases by a more significant amount when $e$ is increased to 0.05.

The all eccentric and aligned fits are in their own region on the plot. For small separations they are almost as long-lived as initially circular runs, but at large separations the differences become greater. Overall, they aren't as well fit by a single exponential relationship as are the sets with only one planet initially eccentric.

Our findings are applicable to near-resonant pairs of exoplanets whose masses and eccentricities have been constrained using transit timing variations (TTVs) observed by the {\it Kepler} spacecraft.  \cite{hadden2016numerical} found an analytical formula for the relative eccentricity of a pair of planets orbiting near first-order mean motion resonances.  At the same time, their individual eccentricities remain poorly constrained, as similarly described by \cite{jontof2016secure}. Our simulations complement those results by demonstrating that systems of planets on moderately eccentric orbits are almost as stable as those on nearly circular orbits if the relative eccentricities are small, as is the case when each planet starts with $e=0.05$ and periapse angles are aligned within  $\sim 45^\circ$.  As a consequence, stability arguments typically cannot be used to rule out moderately large absolute eccentricities.

\section{Conclusions} \label{sec: conclusion}
We investigated the orbital stability of idealized systems of five identical, closely-packed planets that begin uniformly-spaced in terms of mutual Hill radii. In most of the cases considered,  one planet was initially on an eccentric orbit and the other four planets began on circular orbits. As expected, system lifetimes usually increased for larger orbital separations and dropped for larger  initial eccentricity. For a given initial eccentricity, we found that systems approximately obeyed a log-linear relationship between initial separation and close encounter time, in agreement with previous studies of planets starting with circular orbits, but the slopes dropped as the eccentricity increased. Mean motion resonances reduced system stability, but by smaller amounts as initial eccentricities increased.  

Lifetimes of systems with the outer planet initially eccentric were on average similar to those with the inner planet starting with the same eccentricity; giving the specified initial eccentricity to an intermediate planet resulted, on average, in slightly shorter system lifetimes. The dependence of system lifetime on which planet is initially eccentric do not scale with total system angular momentum deficit, suggesting that mean motion resonances dominate over secular resonances in destabilizing the planetary systems studied. 

Studies in which all planets began with an eccentricity $e=0.05$ produced interesting results. If the planetary periapses were selected randomly, system lifetimes were, as expected, far shorter than comparable systems with only one planet initially eccentric. However, if all  of the planets' periapses were initial aligned or nearly so, systems were generally much longer-lived than systems in which only one planet had initial $e=0.05$. For initially fully aligned systems with all planets beginning with $e=0.05$,  planetary spacing required for  survival  $> 10^9$ years is similar to those for which four of the planets begin on circular orbits and the other planet starts with $e=0.01$.

\section{Acknowledgements} \label{sec: acknowledgements}
We thank Nicolas Faber, Eric Feigelson, Sam Hadden, Daniel Jontof-Hutter, Yoram Lithwick, Fred Rasio, David Rice, Jason Steffen, Daniel Tamayo, and Christa Van Laerhoven for useful comments and discussions.

PG was supported by NASA-Ames / Fonds National de la Recherche Luxembourg Grant SP0044174/60048152.
 JJL's work on this project was supported by NASA's Astrophysics Data Analysis Program 16-ADAP16-0034.

\section{References}

\bibliographystyle{apalike}
\bibliography{refs}

\end{document}